\documentclass[a4paper,fleqn]{cas-sc}

\usepackage[authoryear]{natbib}
\usepackage{graphicx} 
\usepackage{float}
\usepackage{algorithm}  
\usepackage{algpseudocode}
\usepackage{color}
\usepackage{setspace}
\usepackage{amsmath}

\newcommand{\vx}{{\mathbf x}}
\newcommand{\vv}{{\mathbf v}}
\newcommand{\va}{{\mathbf a}}
\newcommand{\vu}{{\mathbf u}}

\newcommand{\bx}{\boldsymbol{x}}
\newcommand{\by}{\boldsymbol{y}}
\newcommand{\btheta}{\boldsymbol{\theta}}

\newif\ifnotesw \noteswtrue

\def\tsc#1{\csdef{#1}{\textsc{\lowercase{#1}}\xspace}}
\tsc{WGM}
\tsc{QE}
\tsc{EP}
\tsc{PMS}
\tsc{BEC}
\tsc{DE}

\begin{document}
\let\WriteBookmarks\relax
\def\floatpagepagefraction{1}
\def\textpagefraction{.001}
\shorttitle{Bayesian Learning of Gas Transport in Three-Dimensional Fracture Networks}
\shortauthors{Shi et al.}

\title[mode = title]{Bayesian Learning of Gas Transport in Three-Dimensional Fracture Networks}

\author[1,2]{Yingqi Shi}
\credit{conceptualization, methodology, software, validation, formal analysis, investigation, data curation, writing - original draft, visualization}

\author[1,3]{Donald J.\ Berry} 
\credit{conceptualization, methodology, software, validation, formal analysis, investigation, data curation, writing - original draft, visualization}

\author[1]{John Kath}
\credit{conceptualization, methodology, software, validation, formal analysis, investigation, data curation, writing - original draft, visualization}

\author[1,2]{Shams Lodhy} 
\credit{investigation, data curation, writing - original draft}

\author[1]{An Ly}
\credit{validation, investigation, data curation, writing - original draft, writing - review \& editing, visualization}

\author[1]{Allon G.\ Percus}[orcid=0000-0002-0847-5284]
\credit{conceptualization, methodology, validation, formal analysis, investigation, resources, writing - original draft, writing - review \& editing, visualization, supervision, project administration}
\ead{allon.percus@cgu.edu}

\author[4]{Jeffrey D.\ Hyman}[orcid=0000-0002-4224-2847]
\credit{conceptualization, methodology, software, formal analysis, investigation, resources, data curation, writing - original draft, writing - review \& editing, visualization, supervision, project administration}

\author[5]{Kelly Moran}[orcid=0000-0003-3551-2885]
\credit{conceptualization, formal analysis, resources, writing - review \& editing}

\author[5]{Justin Strait}[orcid=0000-0003-4356-9443]
\credit{conceptualization, methodology, software, formal analysis, resources, resources, writing - review \& editing, supervision}

\author[4]{Matthew R.\ Sweeney}[orcid=0000-0002-5160-4176]
\credit{conceptualization, investigation, resources, data curation}

\author[4]{Hari S.\ Viswanathan}[orcid=0000-0002-1178-9647]
\credit{conceptualization, methodology, writing - review \& editing, supervision, project administration}

\author[4]{Philip H.\ Stauffer}[orcid=0000-0002-6976-221X]
\credit{conceptualization, methodology, resources, writing - review \& editing, supervision, project administration, funding acquisition}

\address[1]{Institute of Mathematical Sciences, Claremont Graduate University, Claremont, CA 91711, USA}
\address[2]{Drucker School of Management, Claremont Graduate University, Claremont, CA 91711, USA}
\address[3]{Center for Information Systems and Technology, Claremont Graduate University, Claremont, CA 91711, USA}
\address[4]{Computational Earth Science Group, Earth and Environmental Sciences Division, Los Alamos National Laboratory, Los Alamos, NM, 87545, USA} 
\address[5]{Statistical Sciences Group, Computer, Computational, and Statistical Sciences Division, Los Alamos National Laboratory, Los Alamos, NM, 87545, USA}


\begin{abstract}
Modeling gas flow through fractures of subsurface rock is a particularly challenging problem because of the heterogeneous nature of the material. High-fidelity simulations using discrete fracture network (DFN) models are one methodology for predicting gas particle breakthrough times at the surface, but are computationally demanding. We propose a Bayesian machine learning method that serves as an efficient surrogate model, or emulator, for these three-dimensional DFN simulations. Our model trains on a small quantity of simulation data and, using a graph/path-based decomposition of the fracture network, rapidly predicts quantiles of the breakthrough time distribution.  The approach, based on Gaussian Process Regression (GPR), outputs predictions that are within 20--30\% of high-fidelity DFN simulation results.  Unlike previously proposed methods, it also provides uncertainty quantification, outputting confidence intervals that are essential given the uncertainty inherent in subsurface modeling.  Our trained model runs within a fraction of a second, which is considerably faster than other methods with comparable accuracy and multiple orders of magnitude faster than high-fidelity simulations.
\end{abstract}


\maketitle 

\printcredits

\doublespacing

\section{Introduction}
\label{sec:intro}

Modeling and predicting the passage of fluids through subsurface fracture networks is a fundamental and ongoing challenge in numerous civil, governmental, and industrial applications. These include the long-term storage of spent nuclear fuel, aquifer management and cleanup, enhanced geothermal energy systems, conventional/unconventional hydrocarbon extraction, geological sequestration of carbon, and the detection of chemical signatures from a clandestine nuclear weapons test~\citep{selroos2002comparison,neuman2005trends,follin2014methodology,sun2014modeling,jenkins2015state,middleton2015shale,hyman2016understanding,bourret2019evaluating,national2020characterization}.
There are a variety of computational methods for modeling gas transport through such systems, and specifically for obtaining the distributions of solute travel times, i.e.,\ the breakthrough curve.
One distinguishing feature between the different modeling approaches is how, and with what level of fidelity, fractures and the surrounding host rock are represented~\citep{hyman2022flow,viswanathan2022from}.

There are continuum methods where the effects of fractures are accounted for using upscaled quantities, e.g.,\ effective permeability and porosity, cf.~\cite{sweeney2020upscaled} for additional references and a more comprehensive description.
However, continuum methods often poorly represent network connectivity, which is believed to be a key geostructural property that controls transport through fractured media~\citep
{hyman2020characterizing,maillot2016connectivity}. 
A result of upscaling the network structure into effective properties is that it becomes challenging to link transport observations with geostructural properties~\citep{kang2020}.
In contrast, there are discrete fracture network (DFN) models where the individual fractures and the interconnected networks they form are explicitly represented. 
While the higher level of fidelity provides a more accurate representation of transport, and the ability to link geostructural attributes with flow and transport observations~\citep{hyman2019emergence,hyman2019linking}, DFN simulations are more computationally expensive than continuum models.
DFN models typically require unstructured computational mesh generation, have a large number of degrees of freedom, and use a computational physics simulator, all of which scale with the spatio-temporal size of the simulations~\citep{ushijima2021multilevel}.
Regardless of the adopted computational methodology, the large uncertainty inherent in subsurface modeling requires that ensembles of realizations are generated to obtain expected transport behavior along with confidence intervals of the breakthrough curves.
This requirement exacerbates the computational demands of subsurface modeling and particularly DFN models. 

A promising alternative to address these computational costs is the use of machine learning (ML) techniques that take geostructural information of the network, e.g., topological, geometric, and hydrological attributes, as inputs to provide estimates of flow and transport observations.
These surrogate models, or emulators, can be trained on a modest amount of high-fidelity simulation data to provide approximations of gas/solute breakthrough times.  
Once trained, ML models generate very rapid predictions, typically within a fraction of a second.  
The predictions may not be as accurate as the direct simulations themselves.
However, owing to their speed, they allow for the study of a much larger sample of fracture networks than would otherwise be possible using conventional computational physics models.
In turn, they can provide a more comprehensive  understanding of the underlying physical phenomena.

Previously, ML approaches have been used to identify primary flow subnetworks \citep{Valera2018,Srinivasan2019,Srinivasan2020Backbone}, though not to generate direct predictions of particle arrival times themselves.  
Furthermore, these techniques do not provide confidence intervals, an essential element in view of the uncertainty associated with subsurface transport problems.
Our work addresses both shortcomings.  We use Gaussian Process Regression (GPR, also known as kriging) to obtain predictions for transport breakthrough times through fracture networks along with uncertainty quantification.  Our results yield increased accuracy and speed compared to other (non-ML) reduced-order modeling of breakthrough times.  Moreover, our method provides robust confidence bounds, blending physics-based modeling with Bayesian inference to generate predictions with greater interpretability.

Specifically, we train our model on breakthrough curves (travel time distributions of particles transported along with the flow) generated using high-fidelity three-dimensional DFN simulations.  These simulations are drawn from an ensemble of semi-generic DFNs, based on commonly observed field site characteristics including hydrological properties. 
We characterize the networks using a graph representation to isolate features associated with simple source-to-target paths in the graph,  which are used as ML model inputs. 
The quantities of interest (QoI) predicted by the model are quantiles of the breakthrough time distribution, notably the 0th (first arrival time), 20th, 50th, 70th, and 90th percentile, as well as the peak arrival time.

One of the benefits of GPR is that, as a Bayesian method, it treats model parameters as random variables rather than deterministic quantities.  Given a prior distribution of the model parameters, GPR uses the training data to update this prior and to generate a posterior distribution.  This posterior distribution establishes rigorous confidence bounds on the predicted QoI, addressing the crucial challenge in subsurface hydrology of quantifying the uncertainty in breakthrough curve predictions.

Our results are summarized as follows.  On DFNs with several hundred fractures, our model trains within seconds and the trained model runs within a fraction of a second, which is multiple orders of magnitude faster than the original simulation.  We obtain predictions on breakthrough times that, depending on the quantile, are within 20--30\% of the values obtained from high-fidelity simulations.  This is a high level of accuracy given the sparsity of experimental data on true subsurface fracture sites and the fact that DFNs themselves provide at best a statistical characterization of such data.  Furthermore, our prediction quality is competitive with the best currently existing reduced-order models for breakthrough time prediction, which obtain similar accuracy but take several times as long to run, due to the need to simulate particle flow.  
Finally, we obtain confidence bounds for our predictions.  
We find that, when predicting the log of the breakthrough time, the width of our 95\% confidence intervals is within 16\% of the predicted value.  Results of this quality can impact numerous application areas, including gas seepage from underground nuclear explosions, natural gas extraction, and detection of methane leakage from wells.

\section{Methods: Flow and Transport in Fracture Networks}
\label{sec:background}

\subsection{Three-Dimensional Discrete Fracture Networks}

For the high-fidelity simulations used to generate our training data set, we adopt a Discrete Fracture Network (DFN) approach to model flow and transport through the fractured rock mass.
In a DFN model, the fractures are represented as a network of intersecting planes whose sizes, shapes, orientations, and hydrological properties are sampled from distributions whose parameters are determined from a site characterization, cf.~\cite{viswanathan2022from} for a comprehensive discussion of DFN modeling approaches.
We use the {\sc dfnWorks} software \citep{Hyman2015}, which provides an end-to-end workflow from network generation to flow and transport simulation. 
We consider an ensemble of semi-generic DFNs generated in a cubic domain with sides of length $L=100$~m. 
The networks are semi-generic in that they do not represent a particular field site, but the characteristics are loosely based on field observations~\citep{bonnet2001scaling}.
Within the DFN, there is one fracture family, with radii drawn from a truncated power law distribution with a decay exponent of 1.5, and lower and upper cutoffs of 1 and 50 meters. 
Fracture centers are uniformly distributed throughout the domain and their orientations follow a uniform distribution projected onto the unit sphere, which mimics a disordered media~\citep{hyman2017dispersion}.
These parameters ensure that there is no single fracture connecting inflow and outflow boundaries. 
Initially, 2000 fractures are placed in the domain.
Isolated fractures and clusters that do not connect the inflow and outflow boundaries are removed because they do not contribute to flow. 
The resulting fracture networks contain around 300 fractures each.
The hydraulic aperture of each fracture is positively correlated with the radius using a power-law relationship, $b = 5\cdot 10^{-5} \sqrt{r}$.
Such correlations are a common modeling assumption for a DFN model~\citep{deDreuzy2002,Frampton2010,Hyman2016,Joyce2014}.
This correlation leads to hydraulic variability as well as the geostructural variability of the network. 

Next, we create a computational mesh representation of networks on which to simulate flow and transport. 
The mesh is a conforming Delaunay triangulation produced using the Feature Rejection algorithm for meshing (FRAM) combined with the near-maximal Poisson sampling method (nMAPS), which are described in \cite{hyman2014conforming,krotz2022maximal}.
We simulate steady-state laminar flow on the dual mesh of the triangulation, the Voronoi control volumes, using the massively parallel flow and transport solver {\sc pflotran} \citep{pflotran-user-ref}.
The distribution of volumetric flow rates and pressure within the network is modeled using Darcy's Law
\begin{equation}\label{eq:darcy}
{\bf q} = -\frac{\kappa}{\mu} \nabla P\, 
\end{equation}
where ${\bf q}$ is the volumetric flux with explicit units of [m$^3$/(m$^2$ s)], $\kappa$ is the permeability [m$^2$], $\mu$ is the fluid viscosity [Pa s], and $P$ is the fluid pressure [Pa].
Simulations are performed with the fluid temperature at 20$^{\circ}$ C, which corresponds to a viscosity of $\mu = 8.9\cdot10^{-4}$ Pa s for the pressure values considered.
Fracture permeability is determined using a local cubic law, $\kappa = b^2/12$.
We drive flow through the domain by applying a pressure difference of 1 MPa across the $x$-axis using Dirichlet conditions on the inflow and outflow boundaries.
Note that Eq.~\eqref{eq:darcy} is linear in the pressure gradient, and thus the value of 1 MPa is arbitrary with respect to the structure of the flow field within the network. 
Neumann, no-flow, boundary conditions are applied along lateral boundaries of the domain as well as along fracture boundaries. 
Gravity is not included in these simulations.
In the DFN model, the matrix surrounding the fractures is impermeable, i.e.,\ there is no interaction between flow within the fractures and the solid matrix.
The numerical solution of Eq.~\eqref{eq:darcy} provides values for the pressure and volumetric flow rates throughout the domain that are used to reconstruct the Eulerian velocity field $\vu(\vx)$ within the DFN using the method provided in \cite{makedonska2015particle,painter2012pathline}.

Transport through the network is simulated using tracer particles that follow pathlines through the velocity field $\vu(\vx)$.
Particles are distributed along the inflow plane of the domain using flux-weighting so that the number of particles at a location is proportional to the incoming volumetric flow rate \citep{hyman2015influence,hyman2021transport,kang2017anomalous,kreft1978on}.
The trajectory $\vx(t;\va)$ of a particle starting at a point ${\bf a}$ at time $t=0$ is given by integrating the advection equation
\begin{equation}\label{eq:trajectory}
\frac{d \vx(t;\mathbf a)}{d t} = \vv_t(t;\va), \qquad  \vx(0;\va) = \va.
\end{equation}
Here, the Lagrangian velocity $\vv_t(t;\mathbf a)$ is given in terms of
the Eulerian velocity $\vu(\vx)$ as $\vv_t(t;\va) = \vu[\vx(t;\va)]$.
The dynamics occurring within intersections are a sub-grid scale process represented using a complete mixing model so that the probability of a particle exiting onto a fracture is proportional to the outgoing volumetric flow rate.
The numerical method implemented is described in \cite{sherman2019characterizing}.

We record the arrival time of each particle to exit the domain having traveled a linear distance of $x_L$, which we denote as $\tau(x_L;\va)$. 
These values are used to construct the relative solute breakthrough (probability density function) at a time $t$ defined as
\begin{equation}\label{eq:pdf}
\psi(t,x_L) =  \frac{1}{M} \int\limits_{\Omega_a} d \va ~ \delta[\tau(x_L,\va) - t],
\end{equation}
where $\delta(t)$ is the Dirac delta function.
We also compute the cumulative solute breakthrough (cumulative density function) as 
\begin{equation}\label{eq:cdf}
\Psi(t,x_L) = \frac{1}{M} \int\limits_{\Omega_a} d \va ~ H[\tau(x_L,\va) - t],
\end{equation}
where $H(t)$ is the Heaviside step function.
In Eqs.~\eqref{eq:pdf} and \eqref{eq:cdf}, $M$ is the total number of particles tracked through the domain, which is set to ten thousand in these simulations. 
A preliminary study showed that these primary observables were not dramatically affected by increasing the number of particles beyond this value.
A single run using 32 processors takes around $\approx$ 1-2 hours depending in the number of fractures in the network.

\subsection{Graph Representation}
\label{sec:graph}

\begin{figure}
\begin{center}
  \includegraphics[width=.5\linewidth]{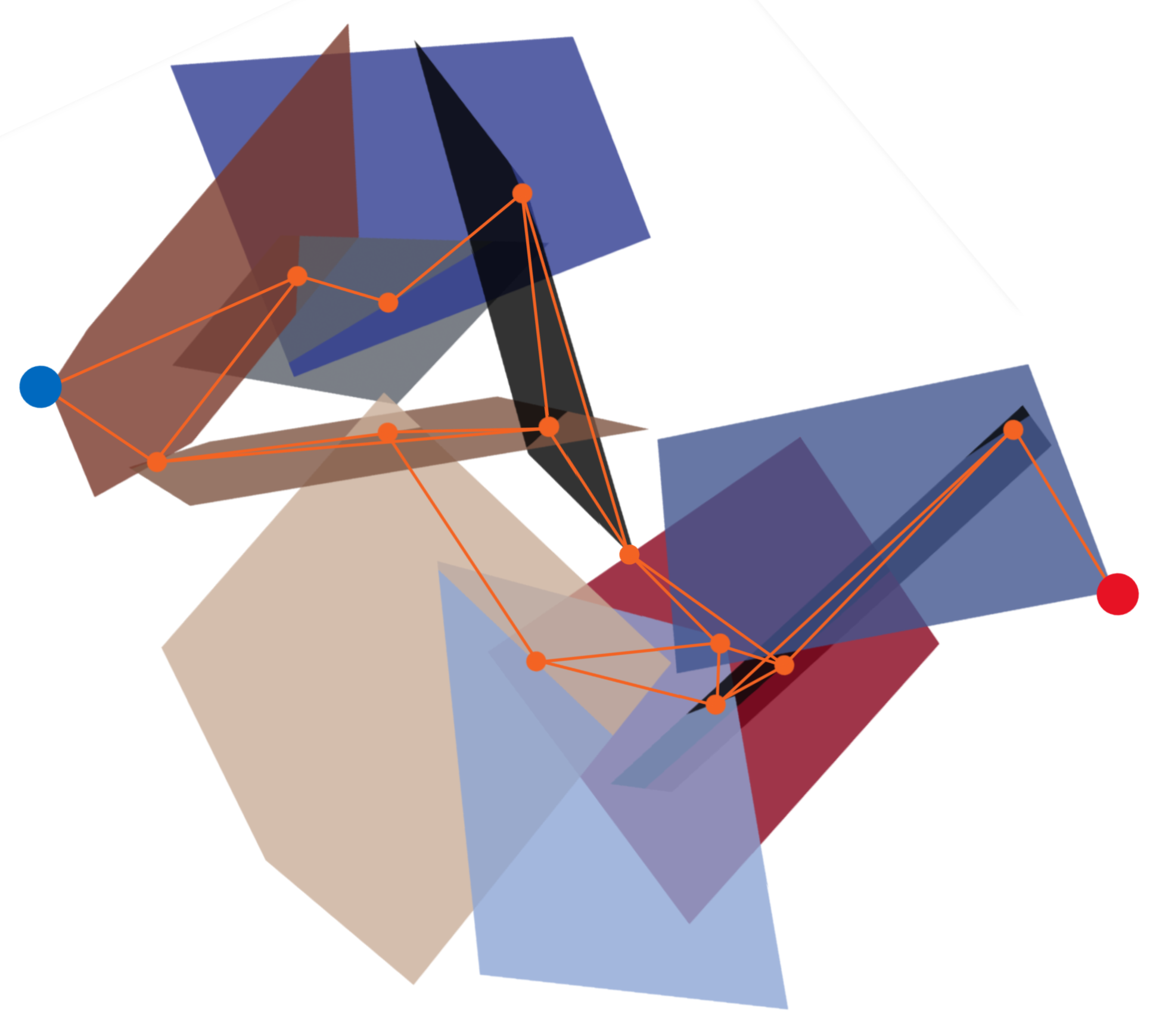}
  \caption{DFN with planes representing fractures, overlaid with its associated intersection graph.  The blue vertex on the left denotes the source (inlet) and the red vertex on the right denotes the target (outlet).\label{fig:dfngraph}}
  \end{center}
\end{figure}

At the center of the DFN methodology is the conceptual model that a set of fractures, which are discrete entities, intersect one another to form a network. 
This conceptual model may be represented as well using the mathematical construct of a graph $G=(V,E)$, which is a tuple consisting of a vertex set $V$ and an edge set $E$ containing pairs of vertices connected by an edge.
There are a variety of graph-based representations of a DFN. 
There is the canonical representation where vertices are fractures and edges are intersections~\citep{andresen2013topology,hope2015topological,huseby1997geometry,hyman2017predictions}, which can be readily populated with particle information to create a flow topology graph~\citep{aldrich2017analysis}, as well as a bipartite graph with one vertex set of fractures and another vertex set of intersections~\citep{Hyman2018Backbone}.  
We adopt a representation where nodes correspond to intersections between fractures $V$, and two vertices are connected by an edge in $E$ if the associated intersections are on the same fracture. 
We refer to this as the \emph{intersection graph} representation.
The precise mathematical formalization and relationship with other representations are presented in \cite{Hyman2018Backbone}.
Figure \ref{fig:dfngraph} shows a small DFN, not one used in the simulations, along with the intersection graph representation as an exposition. 
Additional source and target vertices are added to represent an inlet plane and an outlet plane thereby providing orientation of the graph in topological space.  
Since a fracture can have multiple intersections, it may be represented not only by a single edge, but by a clique (fully connected subgraph) whose edges represent all pairs of intersections.
Additionally, flow and transport can be resolved on the graph representations using the same governing equations provided above, cf.~\cite{dershowitz1999derivation,doolaeghe2020graph,Karra2018} for examples.
For this reason, the intersection graph has been called a pipe or channel network representation of a fracture network~\citep{dershowitz1999derivation}. 
In turn, approximations of mass flux and travel times on edges within the graph can be readily obtained. 

\section{Gaussian Process Regression}
\label{sec:gpr}

Consider a collection of $N$ DFN simulations.
For the $i$th sample, let $G_i$ be its intersection graph representation as defined above and let $y_i$ be its high-fidelity QoI (appropriate quantile of breakthrough time distribution).  Our objective is to learn a surrogate model 
\begin{equation}
\label{eq:surrogate}
y_i = f(x_i),
\end{equation}
where the model input $x_i$ consists of features extracted from the graph $G_i$.
For our surrogate model, we use Gaussian Process Regression (GPR), which is a Bayesian ML approach.  Unlike algorithms that simply attempt to learn exact values for the parameters of a predictive function, Bayesian methods infer a probability distribution over all possible values.  
GPR's ability to work well on training sets of modest size, while offering rigorous confidence bounds on predictions, make it well-suited to our application.  Unlike traditional linear regression models, GPR does not require \emph{a priori} specification of a functional form.  It allows nonlinear fits, and simultaneously provides tractable uncertainty quantification through a closed-form expression.

\subsection{Bayesian Approach}

Consider learning the surrogate model function $f(x)$ in Eq.~\eqref{eq:surrogate}.  In a parametric Bayesian setting, we express the model as $f_{\btheta}(x)$, and assume that while the parameters $\btheta$ are themselves unknown, some
prior distribution $p(\btheta)$ may be assumed.  We then take (generally noisy) training observations of the function, $y=f_{\btheta}(x)+\epsilon$, where $\epsilon$ is a random variable, and use these to update our knowledge of the distribution $p(\btheta)$.  The updated, or posterior, distribution $p(\btheta \mid \by,\bx)$ given the set of measurements $(\bx,\by)$ is computed using Bayes' rule,
\begin{equation}
    p(\btheta \mid \by, \bx)=\frac{p(\by \mid \bx, \btheta)\, p(\btheta)}{p(\by \mid \bx)},
\end{equation}
where $p(\by \mid \bx, \btheta)$ is known as the likelihood and the normalizing factor $p(\by \mid \bx)$ is the marginal likelihood.

The posterior distribution is derived from both the prior distribution and our training observations, which are encoded in the likelihood.  Both the prior and likelihood are often taken to be Gaussian.  Since information about the model parameters is not typically known before seeing training data, it is common to take the prior to be a multivariate unit normal distribution, $\btheta\sim N(0,I)$.  One expects that with increasingly complete training data, the resulting model will be decreasingly sensitive to the exact prior chosen.  The likelihood represents the probability that the model, with inputs $\bx$ and parameters $\btheta$, will give the outputs $\by$, where $\bx$ and $\by$ represent observations in our training set.  We take the likelihood to be a multivariate normal distribution whose mean is given by the model outputs $f_{\btheta}(\bx)$ and whose covariance is given by the assumed covariance of the noise $\epsilon$.

Suppose we now wish to predict the value of the function $f_{\btheta}$ at a new unobserved point of interest, $x^*$, given the dataset of our (noisy) measurements $(\bx,\by)$. The predictive distribution is determined by weighting all possible predictions by their calculated posterior distribution:
\begin{equation}
    p\left(y^{*} \mid x^{*}, \by, \bx\right)=\int_{\btheta} p\left(y^{*} \mid x^{*}, \btheta\right) p(\btheta \mid \by, \bx)\, d \btheta.
\end{equation}

When the prior and likelihood are Gaussian, one can show that the predictive distribution is also Gaussian \citep{Rasmussen2006}. A point prediction can then be found by using its mean, and an uncertainty quantification by using its variance.

\subsection{Gaussian Processes}

While the previous example used a parameterized model, Gaussian processes are a means of performing nonparametric regression.
A Gaussian process $f(x)\sim GP(\mu,\Sigma)$ is a distribution over functions, with mean $\mu(x)$ and covariance $\Sigma(x,x')$, constructed so that any finite set of function evaluations at points $\bx=(x_1,\dots,x_n)$ has a multivariate normal distribution \citep{Rasmussen2006}:
\begin{equation}
    f(\bx)\sim N\left(\mu(\bx),\Sigma(\bx,\bx)\right).
\end{equation}

GPR adopts a Gaussian process as a prior, which it updates using Bayes' rule, conditioning on observed data to generate the posterior distribution over the space of possible functions. Thus, rather than finding a conditional distribution of function parameters $\btheta$, it finds the conditional distribution of the function $f(x)$ itself, directly yielding a predictive distribution for the model.

The mean $\mu(x)$ of the prior Gaussian process is typically set to zero.  However, its covariance $\Sigma(x,x')$, or kernel function, plays an instrumental role in GPR.  It captures the similarity between pairs of input data points and controls the degree to which the output at one point influences the output at another point.  Different covariance functions can capture different patterns of correlation between input data, with the choice of kernel ultimately controlling the smoothness of the function $f(x)$.  A common choice is the squared exponential, or radial basis function (RBF) kernel
\begin{equation}
K(x,x')=\exp\left(-\frac{1}{2\gamma^2} \| x-x' \|^2\right),
\end{equation}
i.e.,\ depending only on the Euclidean distance between $x$ and $x'$.  Note that this choice of $K(x,x')$ is infinitely differentiable, and so Gaussian processes using the RBF kernel are quite smooth in general.  To account for noisy training observations, one can also add a white noise kernel, which is a constant diagonal term equal to the assumed variance $\sigma_\epsilon^2$ of the noise.  This results in the covariance matrix
\begin{equation}
\label{eq:kernel}
\Sigma(\bx,\bx) =
\begin{bmatrix}
K(x_1,x_1) & \cdots & K(x_1,x_n) \\
\vdots & \ddots & \vdots \\
K(x_n,x_1) & \cdots & K(x_n,x_n)
\end{bmatrix}
+ \sigma_\epsilon^2 I.
\end{equation}
The RBF kernel's scale factor $\gamma$ and the noise strength $\sigma_\epsilon$ can be optimized \citep{Rasmussen2006} by maximizing the log marginal likelihood of the observed data,
\begin{equation}
    \max_{\gamma,\sigma_\epsilon}\log p(\by | \bx,\gamma,\sigma_\epsilon) = \max_{\gamma,\sigma_\epsilon}\left(-\frac{1}{2} \left[\by^T\Sigma^{-1}(\bx,\bx)\by + \log\left(\det\Sigma(\bx,\bx)\right) + n \log 2\pi\right]\right).
\end{equation}
Note that in learning values for the hyperparameters via maximum likelihood, we use a convenient frequentist tool within a broader Bayesian context. One could also assign priors to these RBF hyperparameters, and learn their posteriors in a fully Bayesian paradigm.

Using the property that a normal distribution's conjugate prior is itself normally distributed, one finds that GPR's predictive distribution is normal.  For a new unobserved point of interest $x^*$, the posterior mean and variance can be expressed in closed form analytically, as:
\begin{align}
    \mu(x^*) &= \Sigma_*(x^*,\bx)\Sigma^{-1}(\bx,\bx)\by, \\
    \sigma^2(x^*) &= \Sigma_{**}(x^*,x^*)-\Sigma_*(x^*,\bx)\Sigma^{-1}(\bx,\bx)\left[\Sigma_*(x^*,\bx)\right]^T,
\end{align}
where $\Sigma_*(x^*,\bx) = \left[K(x_*,x_1),\dots,K(x_*,x_n)\right]$ and $\Sigma_{**}(x^*,x^*) =K(x_*,x_*)$.

\section{Data}

In this section, we describe our QoI data and the features that we construct as inputs to GPR. 
We train and test the model using a dataset $\mathcal{G}_1$ which contains 100 DFNs. 
For each one of the graphs $G\in\mathcal{G}_1$, a breakthrough curve has been generated using the 3D DFN model. 
In addition to $\mathcal{G}_1$, we have a second dataset $\mathcal{G}_2$ of 100 DFNs, generated using the same statistical generation properties. 
We use this second set only for feature tuning.

\subsection{DFN - Breakthrough Curves}

Using {\sc dfnWorks}, we generate a statistical ensemble of random DFNs, all with the same hydrological properties, for each of the two datasets, and obtain cumulative particle breakthrough densities (Eq.~\eqref{eq:cdf}).
We record quantiles that span much of the breakthrough time distribution, notably 0\% (first arrival time), 20\%, 50\%, 70\%, and 90\%, as well as the peak arrival time.  These form the target quantities of interest (QoI) for our ML model predictions.

\begin{figure}{
  \centerline{\includegraphics[width=0.5\linewidth]{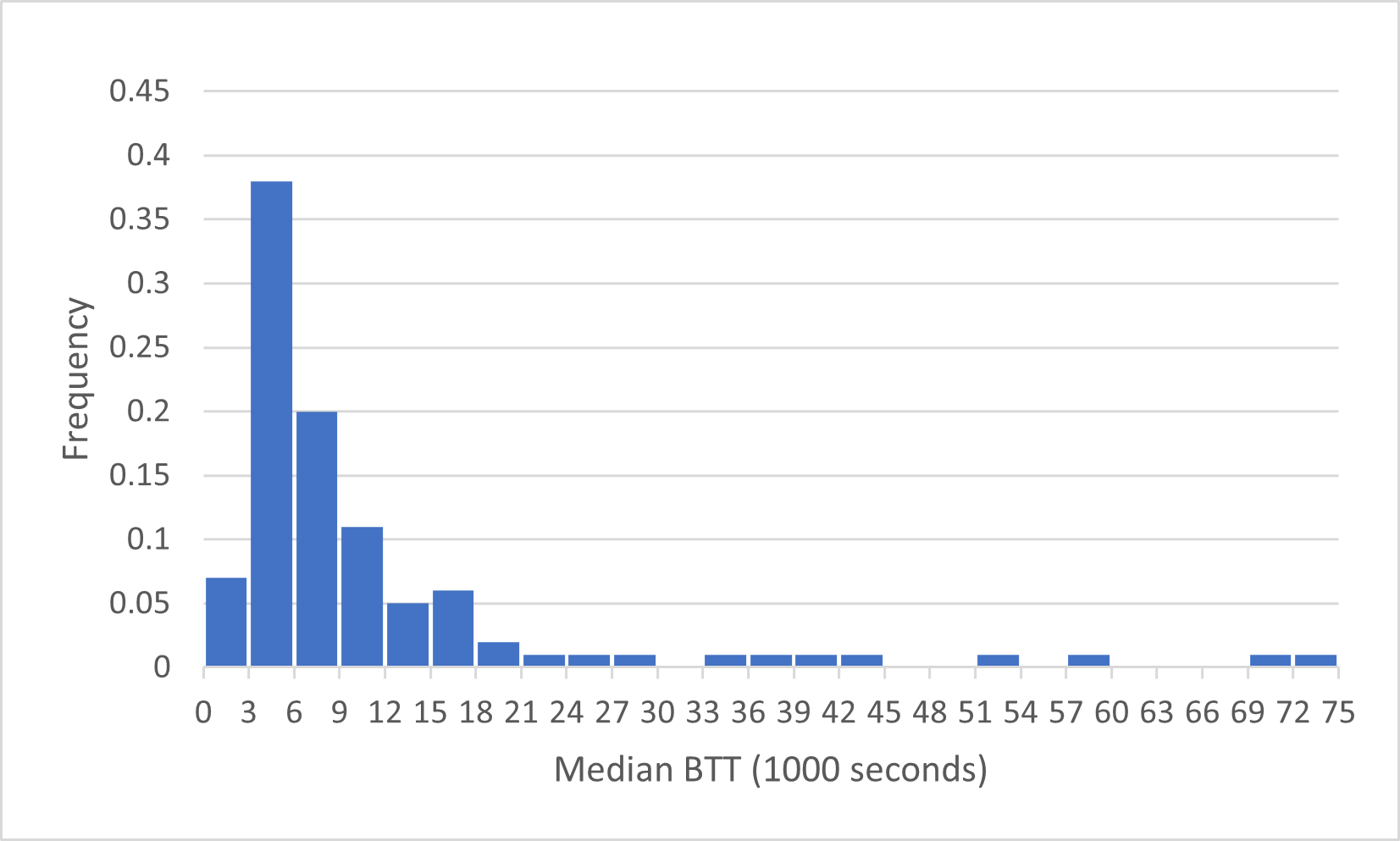}
  \includegraphics[width=0.5\linewidth]{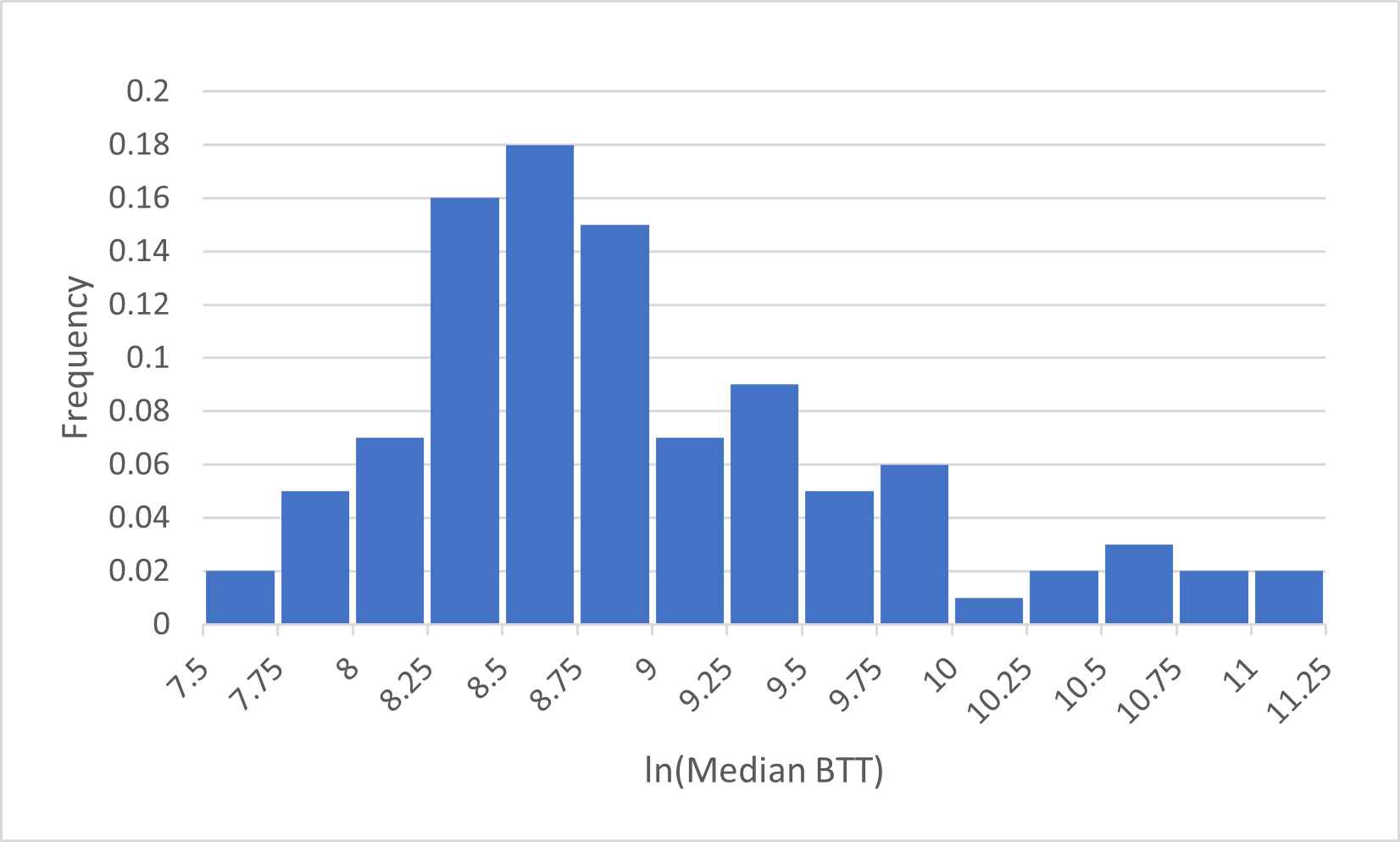}}
  \caption[category]{Median BTT, over the 100 DFNs in the training and testing dataset $\mathcal{G}_1$.  The raw times (left) have a distribution that is closer to log-normal than to normal.  Taking the log of these values (right) transforms the data so that they are closer to normal.\label{Raw-v-log-transformed}}
  }
\end{figure}

GPR functions most effectively when applied to data with a Gaussian distribution. Fig.~\ref{Raw-v-log-transformed} demonstrates that the QoI is heavily skewed, with a median breakthrough time (BTT) displaying a distribution that is closer to log-normal than to normal. We therefore take the log of this quantity, whose distribution is closer to normal (apart from an extended right-hand tail).  We use the log of the BTT quantiles as the output of our model, for training and prediction purposes.

\subsection{Model Input}

Many methods have been developed to extract features from graphs for machine learning \citep{Stamile2021}. They can be grouped into a number of categories such as feature-based methods, shallow embedding methods, regularization methods, and graph neural networks \citep{Chami2021}. As an alternative, we adopt the physics-informed presented by \cite{Srinivasan2020Backbone}.
In order to extract features that describe the flow through the graph, we consider features associated with paths connecting the source to the target node. 
This approach is well-suited to characterizing particle breakthrough times, as it captures attributes that most influence transport across the network.

To form an input data point $x_i$ to our surrogate GPR model $f(x)$ in a given DFN, we take a collection of source-to-target paths, along with the set of features associated with each path.  Given $k$ possible paths and $l$ features for each path, the dimensionality of the model input is $x_i\in\mathbb{R}^{k\times l}$.  
Following \cite{Srinivasan2020Backbone}, we identify the $k$ source-to-target simple paths that are shortest in terms of graph distance on the intersection graph. 
However, rather than taking a fixed value of $k$ as in previous work, we leave it as a tunable quantity in our feature selection (see Section~\ref{sec:feature_tuning} below).

We fix $l=5$, and select the five path-associated features that are described below.  
Our choice of features is motivated primarily by quantities used in the graph flow model presented by \cite{Karra2018}, as discussed in Section~\ref{sec:graph} above.
These differ from the graph centrality-based features that were found to be most impactful in identification of primary flow subnetworks \citep{Valera2018,Srinivasan2019,Srinivasan2020Backbone}, but which we did not find to perform as well for direct prediction of breakthrough times.

\subsubsection{Path Length} 

Path length is the number of distinct fractures involved in the path from the source to target node in the network. In the intersection graph representation (see Fig.~\ref{fig:dfngraph}), this is the number of edges along the path.

\subsubsection{Inverse Permeability: $r_{ij}$}

The permeability $\kappa_{ij}$ of the fracture connecting intersections $i$ and $j$ is a measure of conductance, and its inverse is a measure of resistance. In a resistive network, the resistance of two components in series is the sum of the resistances of the two components.  By analogy with this, we consider the inverse permeability of a fracture,
\begin{equation}
r_{ij} = \frac{1}{\kappa_{ij}},
\end{equation}
to be a quantity whose mean over a sequential path of fractures has physical significance.  We take the value of the inverse permeability feature on a path to be the mean of $r_{ij}$ among all edges (fractures) on that path.

Note that the (arithmetic) mean of the inverse permeabilities is equivalent to the reciprocal of the harmonic mean of permeabilities along the path.  The latter quantity is the equivalent permeability for transporting, across the entire path, the same flux under the same pressure gradient.

\subsubsection{Fracture Length: $L_{ij}$} 

Fracture length is the Euclidean distance $L_{ij}$ between the centroids of intersections $i$ and $j$, representing the distance that a particle travels on the fracture $\{i,j\}$.  We take the value of the fracture length feature on a path to be the mean of $L_{ij}$ among all edges (fractures) on that path.

\subsubsection{Mass Flux: $q_{ij}$}

The mass flux associated with an edge is denoted as
\begin{equation}
q_{ij} = \frac{\kappa_{ij}}{\mu L_{ij}}(P_i - P_j),
\end{equation}
where $\mu$ is the viscosity of the fracture $\{i,j\}$ and $P_i, P_j$ are pressures at nodes $i$ and $j$, calculated using the method described in \cite{Karra2018}.  We take the value of the mass flux feature on a path to be the median of $q_{ij}$ among all edges (fractures) on that path.

\subsubsection{Travel Time: $t_{ij}$}

Edge travel time is denoted as
\begin{equation}
t_{ij} = \frac{L_{ij}\phi_{ij}}{q_{ij}},
\end{equation}
where $\phi_{ij}$ is the porosity of the fracture $\{i,j\}$ taken to be 1 here indicating a completely open fracture.
 We take the travel time feature on a path to be the median of $t_{ij}$ among all edges (fractures) on that path.

\subsection{Feature tuning}
\label{sec:feature_tuning}
While our GPR model does not require any hyperparameter tuning, there is one tunable quantity in our feature construction: the number of shortest source-to-target paths, $k$.  We find the value of $k$ that optimizes model prediction quality, by evaluating the absolute percent error of the predicted median BTT on the 100 DFNs in our second dataset, $G\in\mathcal{G}_2$, using 10-fold cross-validation.  Implementation details for our GPR predictions are described in the Results section below.  In order to avoid data leakage, we use $\mathcal{G}_2$ only for tuning $k$, and not for any subsequent training or testing.

\begin{figure}{
  \centerline{\includegraphics[width=.8\linewidth]{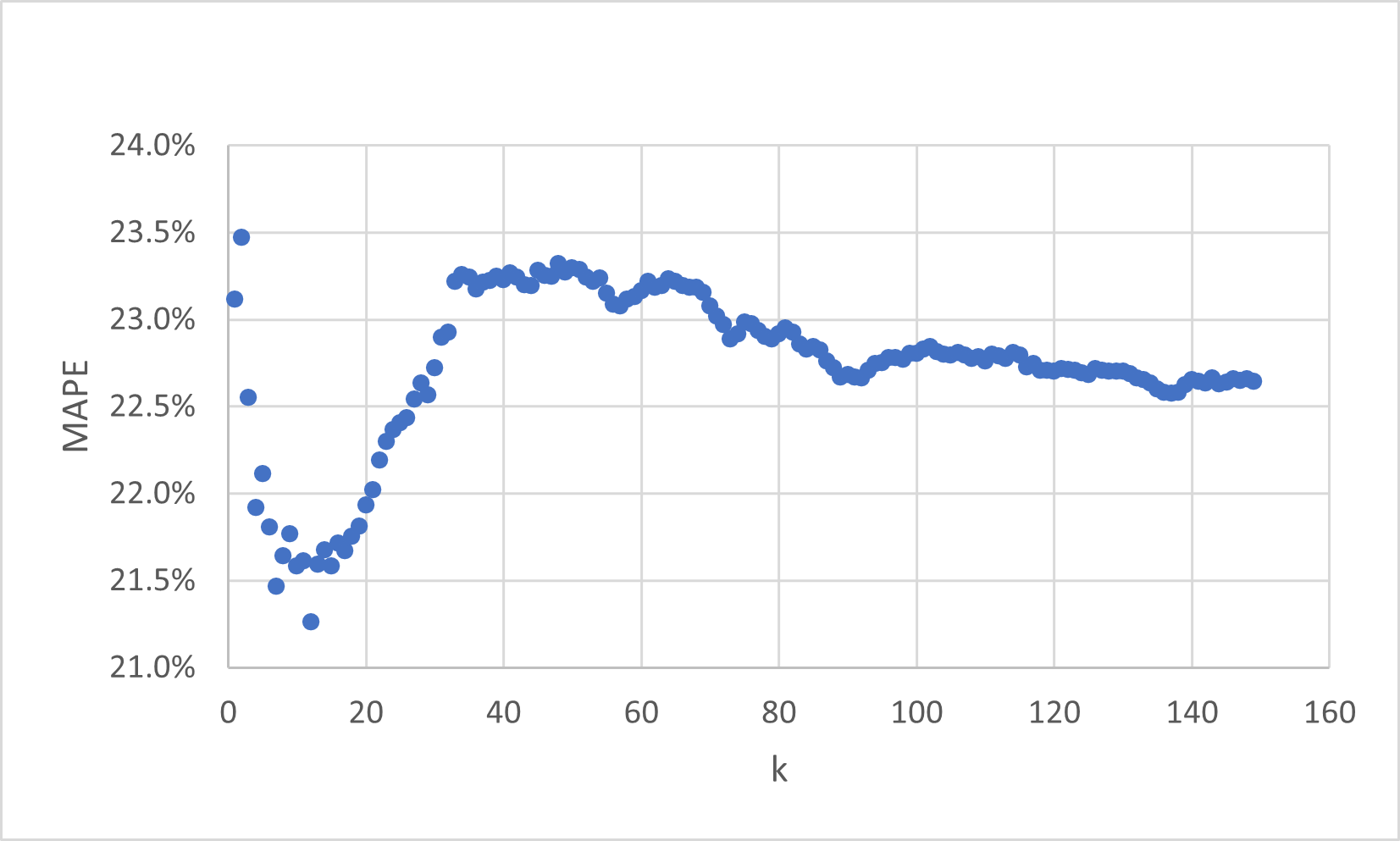}}
  \caption[category]{Mean absolute percent error (MAPE) of predicted median BTT, over the 100 DFNs in the feature tuning dataset $\mathcal{G}_2$, when $k$ shortest source-to-target paths are used in feature construction.\label{Shortest_Paths}}
  }
\end{figure}

Figure~\ref{Shortest_Paths} shows the mean absolute percent error (MAPE), over these 100 DFNs, of the predicted median BTT.  We find that, as a function of $k$, the MAPE has relatively smooth behavior, with values reaching a minimum at smaller $k$ values as more paths are considered.  As $k$ increases further, the inclusion of paths less relevant to particle flow appears to dilute the features, causing MAPE to increase and ultimately stabilize at large $k$.  While the variations are not large (within 2\%), we use the numerical minimum $k=12$ as our optimized value.

Note that, while we optimize $k$ on the basis of the median BTT alone, we use this same tuned value of $k=12$ in the input features for predicting all of our QoI.  It may seem that allowing additional source-to-target paths (larger $k$) could improve performance for larger quantiles of the breakthrough curve such as the 70th and 90th percentile, and that the reverse could hold for smaller quantiles such as the 0th and 20th percentiles.  Empirically, however, we have not found that this gives consistently better predictive performance.

\section{Results}
\label{sec:results}

We implement Gaussian Process Regression (GPR) using the Python {\em scikit-learn} function GaussianProcessRegressor.  We choose an RBF plus white noise kernel, as described in Eq.~\eqref{eq:kernel}, and set all other parameters to default values.

\subsection{Model Predictions}
\label{sec:predictions}

\begin{figure}{
  \centerline{\includegraphics[width=0.5\linewidth]{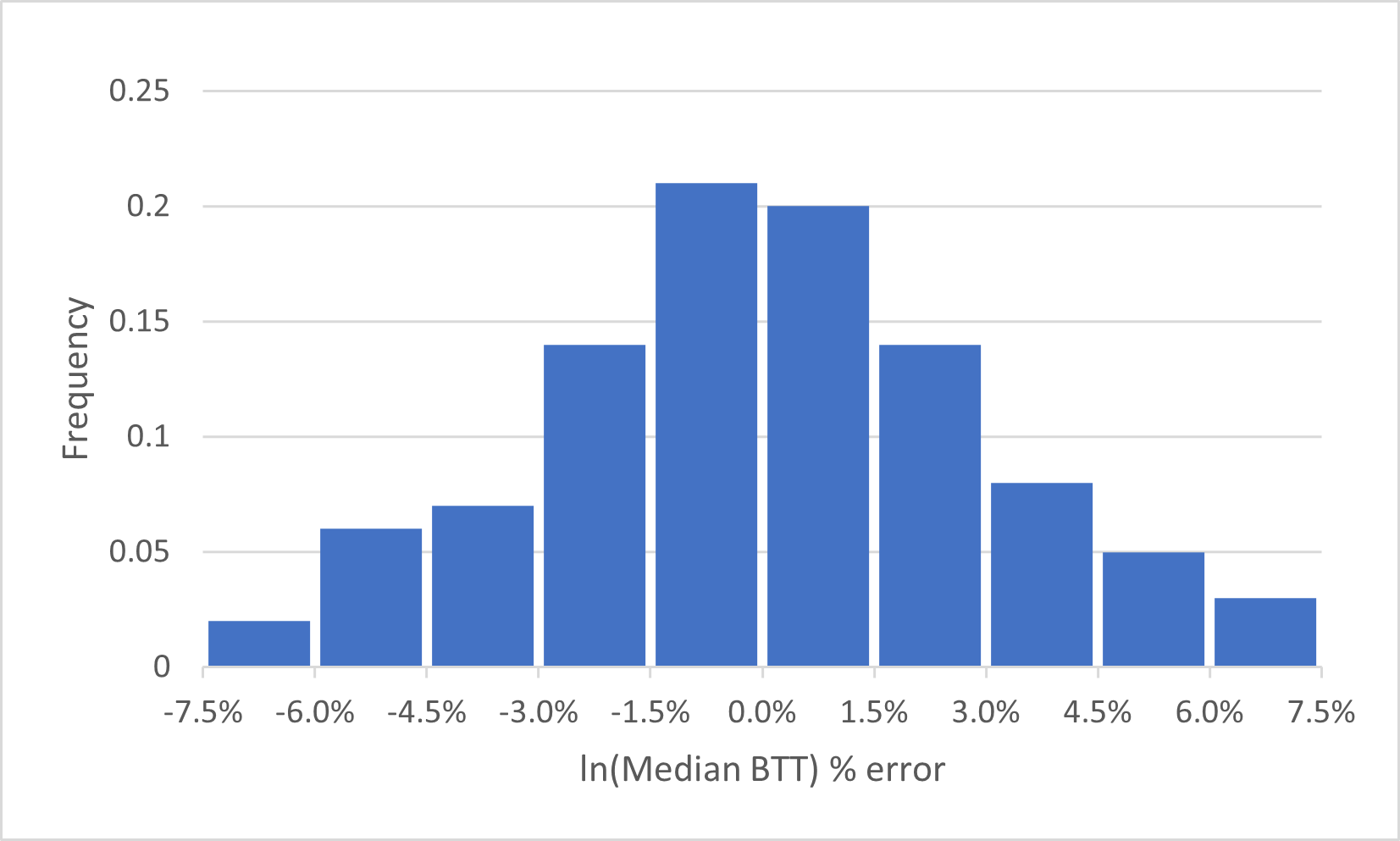}
  \includegraphics[width=0.5\linewidth]{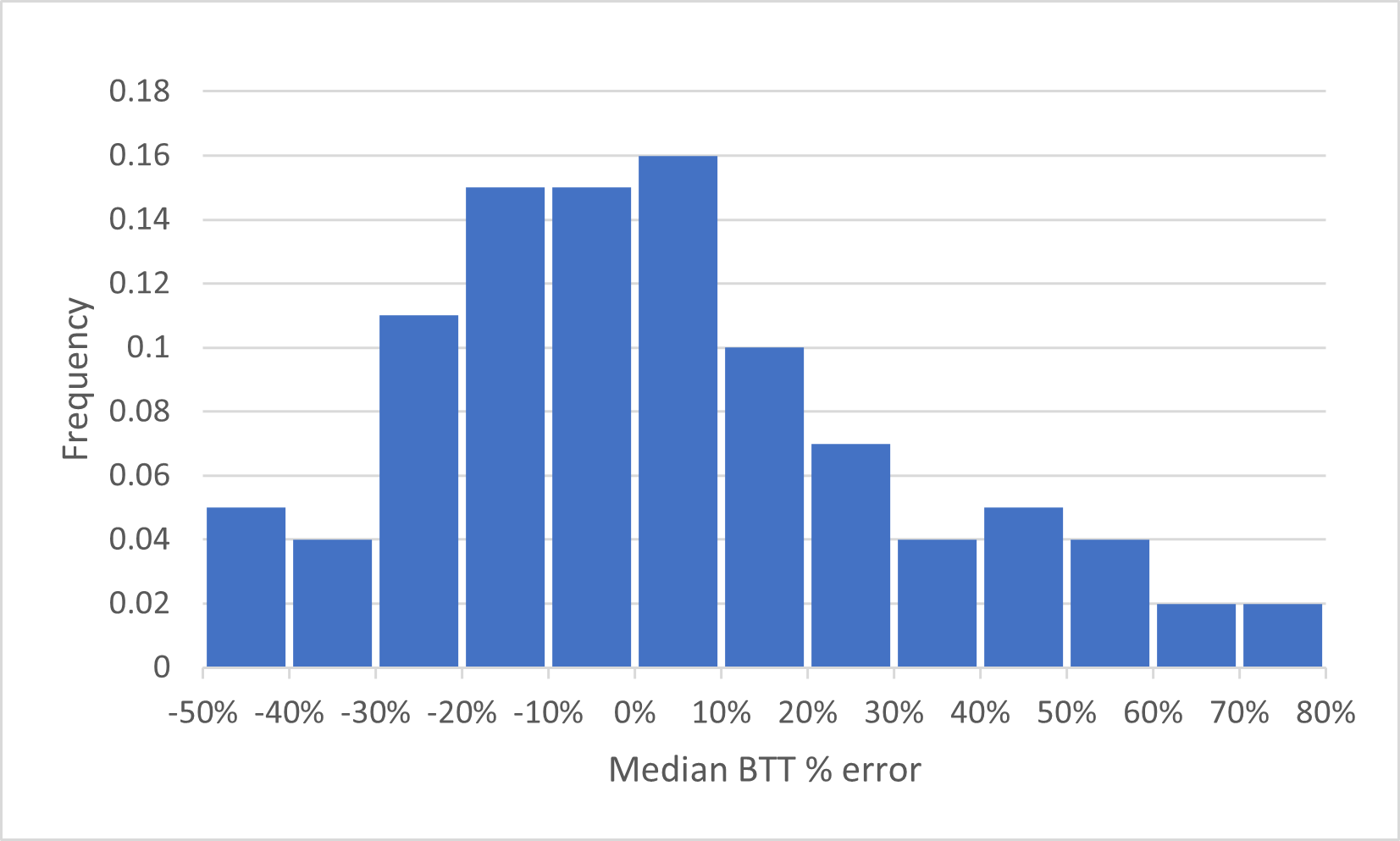}}
  \caption[category]{Percent error of model predictions for median BTT, over the 100 DFNs in the dataset $\mathcal{G}_1$ using 10-fold cross-validation.  Note that distribution is symmetric for predictions of the log of the median BTT (left) and asymmetric when transforming these back to raw median BTT (right).\label{GPR PE}}
  }
\end{figure}

\begin{figure}
  \centerline{\includegraphics[width=.8\linewidth]{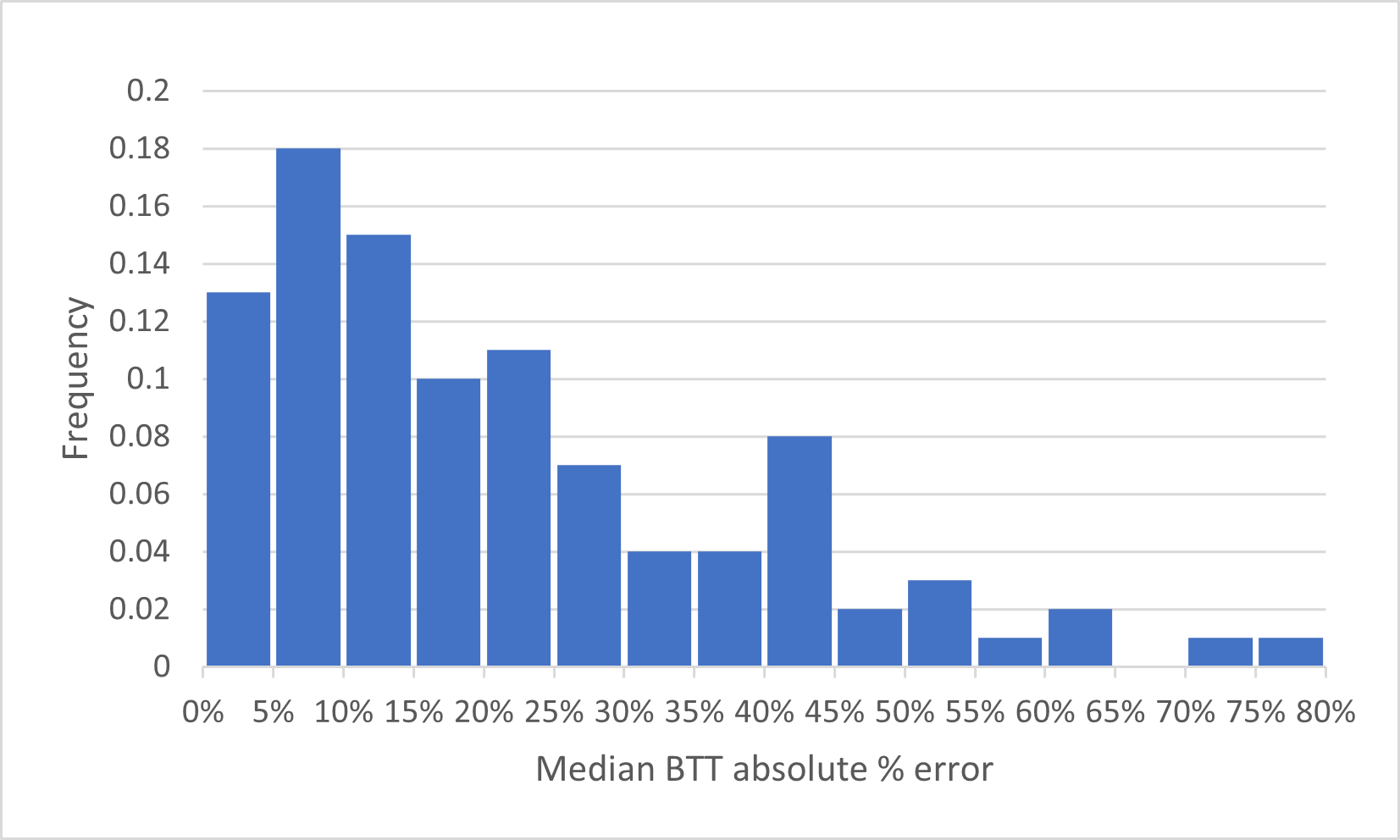}}
  \caption[category]{Absolute percent error of model predictions for median BTT, over the 100 DFNs in the dataset $\mathcal{G}_1$ using 10-fold cross-validation.\label{GPR Absolute PE}}
\end{figure}

We train and test the GPR model on the 100 DFNs in the first dataset $\mathcal{G}_1$ using 10-fold cross-validation.  Recall that our QoI are quantiles of these breakthrough curves, including the 0th (first arrival time), 20th, 50th, 70th, and 90th percentile, as well as the peak arrival time.  Figure~\ref{GPR PE} shows the percent errors of the model predictions of the 50th percentile (median BTT) over this dataset.  Inputs to GPR are standardized, resulting in a symmetric distribution (left) of percent errors in the logarithmic quantity that the model outputs.  This distribution becomes asymmetric (right) when exponentiating the model output to recover one of our QoI, the raw median BTT.  The model predicts this quantity with an error between -47\% and 77\%.
Figure~\ref{GPR Absolute PE} shows the corresponding {\em absolute\/} percent error in the predicted median BTT. For a majority of DFNs, the error is less than 20\%, and exceeds 50\% in only 8 out of 100 cases.

Finally, Table~\ref{mape} expands our comparisons to the other BTT quantiles and peak BTT, showing the mean absolute percent error (MAPE) of model predictions.  Values range from approximately 20\% (for 0th percentile or first arrival time) to 30\% (for 90th percentile).  These numerical results are broadly comparable to the most accurate reduced-order model for BTT prediction, namely the graph flow model of \cite{Karra2018} discussed in Section~\ref{sec:graph}, once the bias-correcting postprocessing step has been applied to that method.  However, our GPR model, once trained, runs in a fraction of a second rather than the seconds required for simulating transport in the graph model.

\begin{table}
\centering
\caption{Mean Absolute Percent Error (MAPE) of model predictions for different BTT quantiles and peak BTT, over the 100 DFNs in the dataset $\mathcal{G}_1$ using 10-fold cross-validation.  0th percentile represents first arrival time of gas particles.\label{mape}}
\begin{tabular}{|l||c|} 
 \hline
 BTT Percentile & MAPE \\ 
 \hline 
 \hline
0th & 20.53\% \\
\hline
20th & 22.53\% \\
\hline
50th & 21.80\% \\
\hline
70th & 23.29\% \\
\hline
90th & 30.24\% \\
\hline
Peak & 26.93\% \\
\hline
\end{tabular} 
\end{table}

\subsection{Uncertainty Quantification}

\begin{figure}{
  \centerline{\includegraphics[width=.8\linewidth]{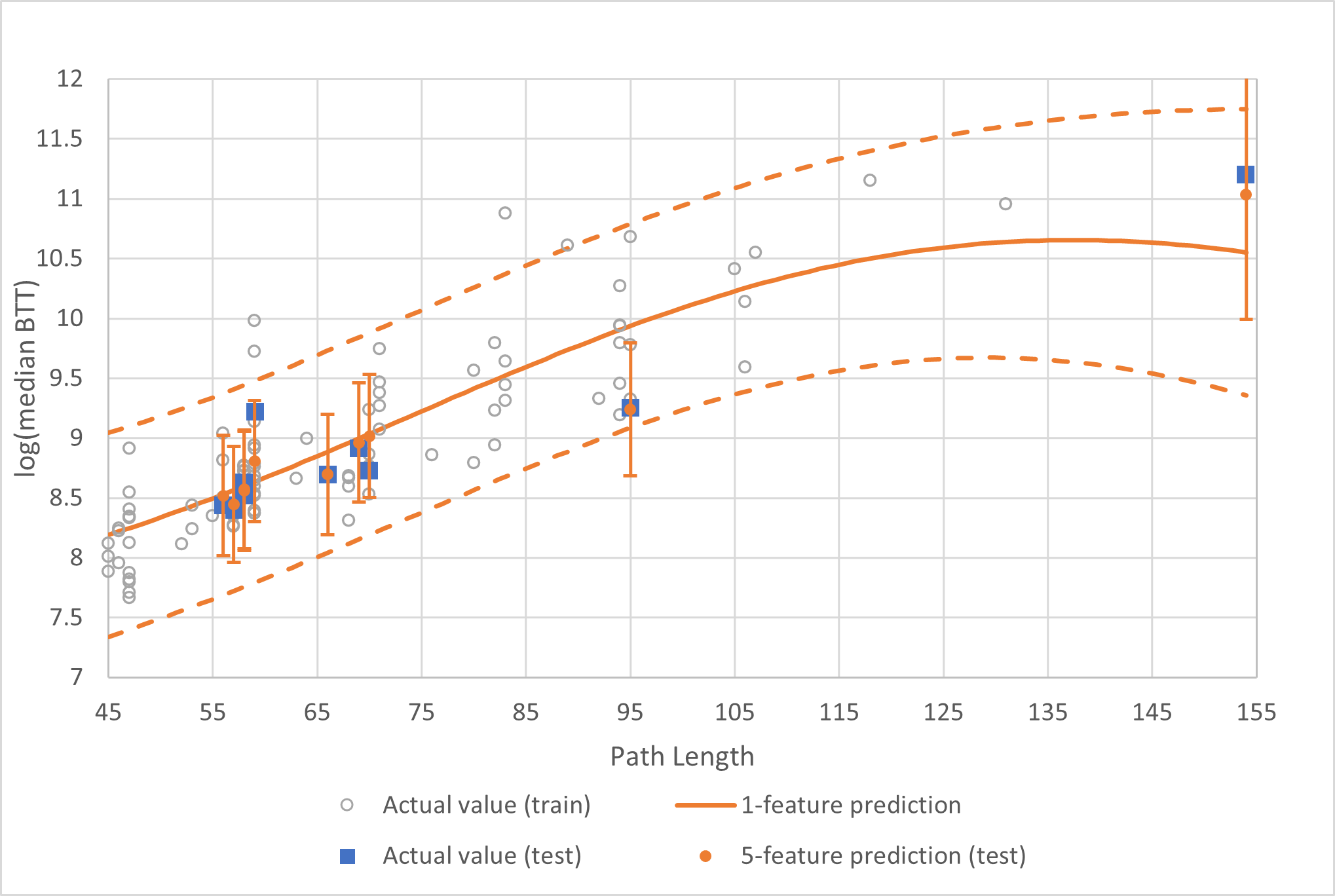}}
  \caption[category]{
  Results from simplified GPR model trained only on 1 feature (path length), along with 95\% confidence bounds, training data points, test data points, and 5-feature predictions (with full GPR model) on test data points.  Grey circles show 90 DFNs from dataset $\mathcal{G}_1$ used to train GPR model.  Orange solid line shows 1-feature predictions varying smoothly with model input (path length).  Orange dashed lines show associated 95\% confidence bounds.  Blue squares show 10 DFNs from dataset $\mathcal{G}_1$ used to test GPR model. Orange points show 5-feature predictions on test DFNs, along with bars showing 95\% confidence bounds for those predictions. \label{1-feature prediction}}
  }
\end{figure}

A further advantage of the GPR model is that it outputs not only a value but a predictive distribution, which provides rigorous confidence bounds on predictions.  In order to illustrate the effectiveness of the method on our problem, we first show the results on a simplified version of it.  Figure~\ref{1-feature prediction} displays predictions for the log of the median BTT when the model is trained using data described by only one feature, the path length.  This allows us to represent, on a 2D plot, a smooth curve showing the model output as that one feature is varied, along with the predictive interval delimited by 95\% confidence bounds (dashed lines).  In this example, the training data consist of 90 fixed DFNs from dataset $\mathcal{G}_1$, denoted by grey circles, and the test data consist of the remaining 10 DFNs, denoted by blue squares.  Finally, orange points represent predictions given on the test data by the full 5-feature model, with error bar whiskers representing 95\% confidence bounds for those predictions.

A few properties are notable in the results in Figure~\ref{1-feature prediction}.  Qualitatively, the predictive interval widens as expected at large path lengths, where training data are sparser and the model is less confident in its predictions.  Quantitatively, the model correctly learns confidence intervals from the training data, with approximately 95\% of those data points (grey) contained within the 95\% confidence bounds for the 1-feature prediction.  This effect is validated on test data, again with approximately 95\% of those data points (blue) contained within the 95\% confidence bounds for the 5-feature prediction.  Not surprisingly, these confidence bounds are considerably tighter than those for the 1-feature prediction, demonstrating the decrease in uncertainty from including the four additional features.

Table~\ref{confidence} summarizes uncertainty quantification results for all of our QoI.  At a given quantile, and for each DFN, we consider the log BTT predicted by the model along with its associated 95\% confidence interval.  We express the width of this confidence interval as a percentage of the log BTT, and average that quantity over all 100 DFNs in the dataset $\mathcal{G}_1$.  This relative confidence interval width varies from 12.30\% of the predicted log BTT (for 50th percentile, i.e.,\ median) to 15.43\% of the predicted log BTT (for peak breakthrough).  Note, though, that relative confidence intervals widths are considerably larger (and asymmetric) when expressed as a percentage of the raw BTT, as one might expect given the MAPE values of 20\% to 30\%.

Table~\ref{confidence} also shows the capture rate: the fraction of actual values that are within the 95\% confidence bounds of the prediction.  These are all close to 0.95, consistent with the model correctly learning confidence intervals over all quantiles.

\begin{table}
\centering
\caption{\% Width of CI represents relative width of predicted confidence interval, expressed as percentage of predicted log BTT, and averaged over the 100 DFNs in the dataset $\mathcal{G}_1$ using 10-fold cross-validation.  Capture rate is fraction of actual values that are within 95\% confidence bounds of the prediction.
Results are shown for different BTT quantiles and peak BTT.\label{confidence}}

\begin{tabular}{|l||c|c|} 
 \hline
 BTT Percentile & \% Width of CI & Capture rate \\ 
 \hline 
 \hline
0th & 12.76\% & 0.92\\
\hline
20th & 12.93\% & 0.97\\
\hline
50th & 12.30\% & 0.93\\
\hline
70th & 13.11\% & 0.95\\
\hline
90th & 14.84\% & 0.93\\
\hline
Peak & 15.43\% & 0.96\\
\hline
\end{tabular} 
\end{table}

\section{Conclusions}
\label{sec:conclusions}

Predicting gas breakthrough times in fracture networks is a crucial scientific challenge, for which high-fidelity DFN simulation methods are computationally expensive. In this paper, we have introduced a Bayesian machine learning approach that uses Gaussian Process Regression (GPR) as an emulator for these simulations, providing rapid predictions of quantiles of the breakthrough time distribution along with confidence intervals that are consistent with the simulation data. Our model is trained on a modest amount of high-fidelity simulation data, using a combination of topological and geological attributes of the DFNs as features.  It generates results within a fraction of a second, making it an attractive choice for modeling subsurface systems where uncertainties in hydrological properties require a large-scale ensemble approach. Our results show that the model's predictions are within 20-30\% of those from high-fidelity simulations.  This is competitive in terms of accuracy with the best currently existing reduced-order models for breakthrough time prediction, which run several times slower than our model, and of sufficient quality to impact application areas ranging from extraction of hydrocarbons to detection of underground nuclear explosions.  Moreover, our Bayesian approach provides rigorous uncertainty quantification, offering analytically tractable confidence bounds.  These are invaluable given the modeling uncertainties in subsurface hydrology, and essential to interpreting model predictions.

It is possible that some improvements to our input features could result in improvements in prediction accuracy. For both the mass flux and travel time, instead of using the median value along a path, we can set a different quantile as a tunable quantity in our feature construction.  Currently, our only such tunable quantity is $k$, the number of shortest source-to-target paths used in feature construction.  In principle, there is no difficulty in optimizing multiple quantities in feature selection, using a grid search, over the same separate dataset ${\mathcal G}_2$ that we now use to tune $k$ alone.  One might also consider the reverse approach of eliminating the mass flux and travel time features altogether, sacrificing accuracy for speed, as the calculation of pressures (required for mass flux and travel time) dominates both the training and the prediction time in our current model.  For networks of several hundred fractures, eliminating that bottleneck could speed up the process by several orders of magnitude.  The results in Figure~\ref{1-feature prediction}, obtained from running GPR with only the path length feature, suggest that such an approach may be viable.

Finally, work is currently underway on generalizing our GPR method to a multi-fidelity setting.  We have seen that graph-based algorithms can be used to supply low-fidelity BTT predictions.  While such methods are slower than GPR, they are orders of magnitude faster than high-fidelity DFN simulations and could therefore serve as a source of less expensive (and consequently more abundant) training data.  Preliminary results suggest that, with a multi-fidelity approach based on either an autoregressive training scheme or a linear Gaussian network, one can obtain results of comparable quality to those in this paper while using far less high-fidelity training data.  This research is ongoing.

\section{Acknowledgments}
The authors would like to acknowledge support from Los Alamos National Laboratory LDRD Award 20220019DR.
Los Alamos National Laboratory is operated by Triad National Security, LLC, for the National Nuclear Security Administration of U.S. Department of Energy (Contract No. 89233218CNA000001).
This work was conducted within Claremont Graduate University's Engineering and Computational Mathematics Clinic program.

\section*{Computer code availability}

Name: {\sc dfnWorks}

Contact: Jeffrey D.\ Hyman, jhyman@lanl.gov

No hardware requirement

Program language: Python 3

Software requirement: Docker (optional)

The software is open source and can be obtained at \url{https://github.com/lanl/dfnWorks}.  A docker container is also available at \url{https://hub.docker.com/r/ees16/dfnworks}.

\bibliographystyle{cas-model2-names}
\bibliography{bibliography} 

\end{document}